\documentclass[review,5p,times,onecolumn,preprint]{elsarticle}
\usepackage{graphicx}
\usepackage{amssymb}
\usepackage{lineno}
\usepackage[version=3]{mhchem} 
\usepackage {xcolor}

\biboptions{numbers,sort&compress}

\journal{Materials Chemistry and Physics}

\begin{document}

\begin{frontmatter}

\title{Self-assembly of a binary mixture of iron oxide nanoparticles in Langmuir film: X-ray scattering study}

\author[label1,label5]{V. Ukleev}
\address[label1]{National Research Centre "Kurchatov Institute" B. P. Konstantinov Petersburg Nuclear Physics Institute, 188300 Gatchina, Russia}
\ead{ukleev@lns.pnpi.spb.ru}
\fntext[label5]{Present address: RIKEN Center for Emergent Matter Science (CEMS), Wako 351-0198, Japan}
\author[label2]{A. Khassanov}
\address[label2]{Institute of Polymer Materials of the Department of Materials Science Friedrich-Alexander University Erlangen-N\"{u}rnberg, Martensstrasse 7, D-91058 Erlangen, Germany}
\author[label3]{I. Snigireva}
\address[label3]{European Synchrotron Radiation Facility, 71, Avenue des Martyrs, CS40220, F-38043 Grenoble Cedex 9, France}
\author[label3]{O. Konovalov}
\author[label1]{M. Dudnik}
\author[label1]{I. Dubitskiy}
\author[label3,label4]{A. Vorobiev}
\ead{avorobiev@ill.fr}
\address[label4]{Department of Physics and Astronomy, Uppsala University, Box 516, 751 20, Uppsala, Sweden}

\begin{abstract}
In present study we exploited Langmuir technique to produce self-assembled arrays composed of monodisperse iron oxide nanoparticles 10 nm and 20 nm in diameter and of their binary mixture. A combination of in-situ X-ray reflectometry and Grazing-incident small-Angle X-ray scattering was used to obtain in-plane and out-of-plane structure of the arrays directly on the water surface. Surface pressure isotherms and X-ray reflectometry analysis showed that monodisperse 10 nm nanoparticles form a highly ordered monolayer, while 20 nm particles pack in three-dimensional clusters with a short-range (nearest-neighbor) correlations between the particles. In a binary mixture of 10 nm and 20 nm nanoparticles composed in proportion 3:1 the self-assembly process results in a structure where the monolayer of 10 nm particles is perturbed by the larger particles. Non-trivial mixing causes an enlargement of interparticle distance but keeps the symmetry of two-dimensional lattice of smaller nanoparticles. Estimation of the acting interactions and micromagnetic simulation suggest the optimal formation for monodisperse and binary ensembles. 
\end{abstract}
\begin{keyword}
Magnetic nanoparticles \sep X-ray scattering \sep Langmuir film


\end{keyword}

\end{frontmatter}


\section{Introduction}

Discovery of new materials with advanced physical characteristics often leads to a creation of new technologies. Nowadays systems exhibiting properties of self-organization and self-assembly are indispensable part of the development of such technological areas as spintronics, photonics and nanoelectronics. In this context, two-dimensional structures assembled from single-domain magnetic nanoparticles (MNPs) attract considerable attention in view of possible practical applications \cite{1,2,3,4,5,6}. MNPs in combination with nanoparticles of other types are considered as a model structure to design metamaterials and to study fundamental properties of specific nanocrystals in confined geometries \cite{19}. For example, binary mixture of MNPs and semiconductor quantum dots is proposed for creation of novel magneto-optic materials \cite{21,lambert2010langmuir}; proper choice of of MNPs tunes magnetic properties of the resulting compounds due to the proximity effect and magnetic dipole interaction \cite{22}. Recently, a long-range ordering of binary mixtures containing magnetic and non-magnetic nanoparticles was revealed by X-ray scattering \cite{17,18} and local hexagonal close-packed order in binary mixture of Fe$_3$O$_4$ and Co nanoparticles mixture was demonstrated by Tunnel electron microscopy (TEM) \cite{22}.

In present study we have exploited Langmuir technique to create ordered monodisperse and binary arrays of MNPs on water surface. MNPs on water surface mostly interact through magnetic dipole-dipole and Van der Waals forces, which depend on the nanoparticle volume and interparticle distance. Aspects of this dependence in ensembles of monodisperse MNPs of different size were studied in details elsewhere \cite{A}. It was shown, that 10 nm iron oxide MNPs tend to form long-range ordered hexagonal monolayers, while 20 nm nanoparticles agglomerate in three-dimensional aggregates. Using the same Langmuir technique we prepared single-component (MD) and binary (BM) mixtures of monodisperse iron oxide MNPs of 10 nm and 20 nm.

To investigate internal organization of the resulting structures we used a powerful combination of X-ray reflectometry (XRR) and Grazing-incident small-angle X-Ray scattering (GISAXS) which allowed to make the complete structural examination of the nanoparticles ordering across the layer and in its plane at the different steps of Langmuir-Blodgett film formation.

We discuss the results of the 20 nm and 10 nm MNPs layer assembling in terms of the interplay of magnetic dipole-dipole and Van der Waals interactions and micromagnetic calculations in order to compare demagnetization energies of single component and binary mixtures of MNPs.

\section{Samples}

Highly monodisperse spherical iron oxide MNPs were purchased from Ocean NanoTech, USA shortly before the experiments. The mean diameter of the particles was obtained by TEM measurements performed by the manufacturer to 10 nm and 20 nm with the size tolerance of 2.5~nm. To prevent coagulation, nanoparticles were stabilized by a monolayer of oleic acid (C$_{18}$H$_{33}$COOH) with corresponding thickness of 2 nm and dispersed in chloroform. In work \cite{A} it has been shown that monodisperse 10nm MNPs form perfect monolayers on water surface. This confirms that no prior ordering or agglomeration happen in their bulk solution before the deposition. Such solutions being stored at ambient conditions remain unchanged over a few years. At the same time, some short-range preordering is not excluded in the bulk solution of 20 nm MNPs due to much stronger dipole-dipole interaction. However, from analysis performed in \cite{A} one can conclude that agglomeration in its essential part starts on water surface and develops further on a solid surface after deposition, but not in the bulk solution. Samples containing 10 nm and 20 nm MNPs are designated as MD10 and MD20, respectively. The binary sample designated as BM10$_3$20 was prepared by mixing original chloroform solutions of 10 nm and 20 nm MNPs so that the particle number ratio was 3:1. This number was chosen on a geometrical assumption for the most compact packing of the plane with the spheres of these two sizes \cite{kennedy2006compact}.

\section{Experiment}

X-ray scattering measurements were performed at ID10 beamline at European Synchrotron Radiation Facility (ESRF, Grenoble, France), which is especially designed for studies on liquid surfaces using the grazing incidence X-ray scattering techniques \cite{smilgies2005troika}. Detailed description of these surface-sensitive techniques can be found elsewhere \cite{tolan1999x,11}. Photons with wavelength $\lambda=1.54$ \AA~were used. XRR data was acquired by one-dimensional position-sensitive detector (PSD) Vantec. In GISAXS experiment the scattering geometry was set by the grazing incident angle $\alpha_i$ and two scattering angles $\alpha_f$ and $\varphi$, which determined the wave vector transfer components $Q_z$ and $Q_y$, which are perpendicular and parallel to the sample plane respectively. Scattered intensity in the ($\alpha_f$, $\varphi$) plane was acquired by two-dimensional PSD Pilatus 300K (487 $\times$ 619 pixels with a pixel size 172 $\mu m^2$). A beamstop was used for the detector protection in the position of direct and specularly reflected beams.

The Langmuir MNP films were prepared in a custom-designed Langmuir trough, installed directly on the sample goniometer with use of an active anti-vibration device Halcyonics MOD2-S. All samples were prepared at room temperature and in the same way, by casting of small droplets from a micro syringe on different parts of the surface area. All sample solutions was highly diluted (1.2 -- 1.3 mg/mL) to ensure uniform distribution of the particles over the surface and to avoid their local agglomeration during evaporation of the solvent. After the spreading, the samples were left along for 15 min to ensure a complete evaporation of the solvent. Then, the trough was sealed and filled with humid helium to minimize X-ray scattering on air and to compensate for evaporation of water. Surface pressure-area isotherms were measured during compression of the layers provided by the moving barrier. The surface pressure was controlled using of a Wilhelmy plate made of Watmann paper and a microbalance (model PS4, Nima Technology, Ltd).

\subsection{Pressure-area isotherms}

\begin{figure}
\includegraphics[width=8cm]{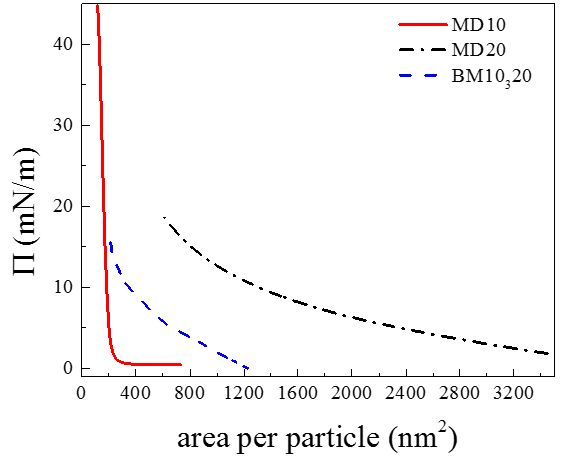}
\caption{Surface pressure isotherms obtained for MD10 (solid red line), MD20 (black dot-dashed line) and BM10$_3$20 (blue dashed line).}
\label{Fig0}
\end{figure}

Using terminology proposed by Harkins \cite{20} for fatty acid monolayers, an isotherm for MD10 sample (Fig.\ref{Fig0}) consists of three regions corresponding to mixed liquid-expanded and gaseous state (LE+G), liquid-condensed state (LC) and solid state (S). Interestingly, transition from the LE+G state to the S state happens almost instantly at the moment when a single particle occupies a surface area equal to its projection on the surface of water. Further compression of the S state to a higher pressure $\Pi = 50$ mN/m does not lead to a transformation of the monolayer in to bi- or multilayers, what is evident from the constant slope of the curve.

An isotherm obtained for the MD20 sample differs drastically from that of obtained for MD10. Pressure starts to increase immediately upon the compression. A smooth transition from the LC state to the S state takes place at only at highest achievable value $\Pi = 14-16$ mN/m, which is much lower than for MD10. Thus almost entire isotherm corresponds to the LC state, which exists already at a very large surface available for a single nanoparticle. We consider this as an indication of a presence of strong long-range interactions between the particles.

An isotherm of BM10$_3$20 sample shows intermediate behavior between MD10 and MD20 isotherms. As in the case of MD20, a monotonic increment upon the pressure $\Pi = 10$ mN/m takes place characterizing the sample as being in the LC state. Further compression leads to a transition to the S state. Notably, the transition starts roughly at the same area values as for MD10 system.

\subsection{X-ray Reflectometry}

\begin{figure}
\includegraphics[width=8cm]{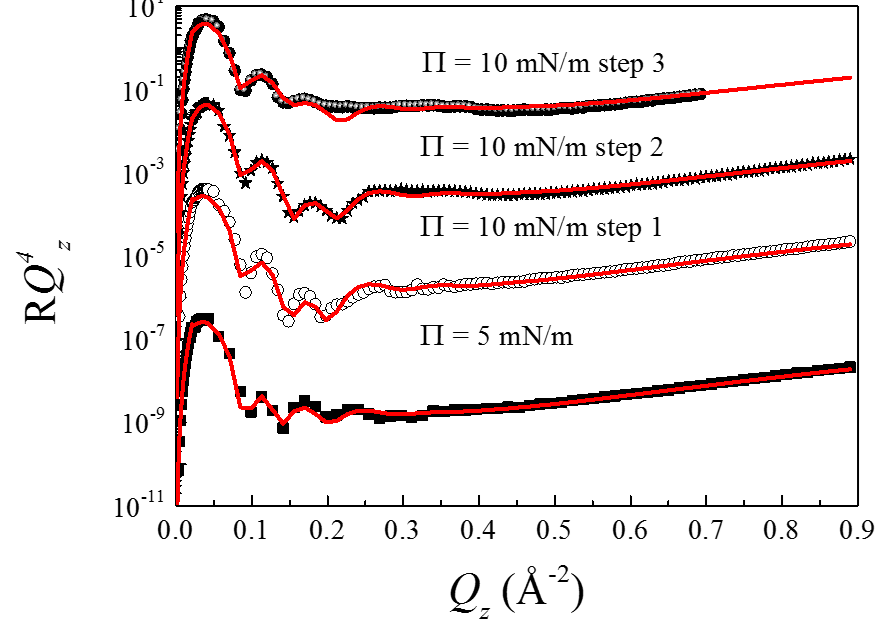}
\caption{XRR experimental data (symbols) and fitted models (solid lines) obtained for MD10 sample at pressure $\Pi=5$ mN/m and $\Pi=10$ mN/m. Measurements at higher pressure were taken three times in a sequential order which is denoted here as step 1, step 2 and step 3}
\label{Fig1}
\end{figure}

\begin{figure}
\includegraphics[width=8cm]{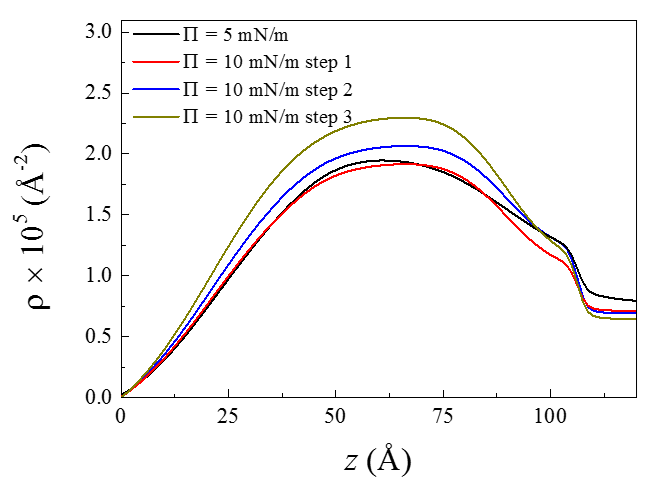}
\caption{SLD profiles of MD10 sample at different steps of layer formation at pressure.}
\label{Fig2}
\end{figure}

\begin{figure}
\includegraphics[width=8cm]{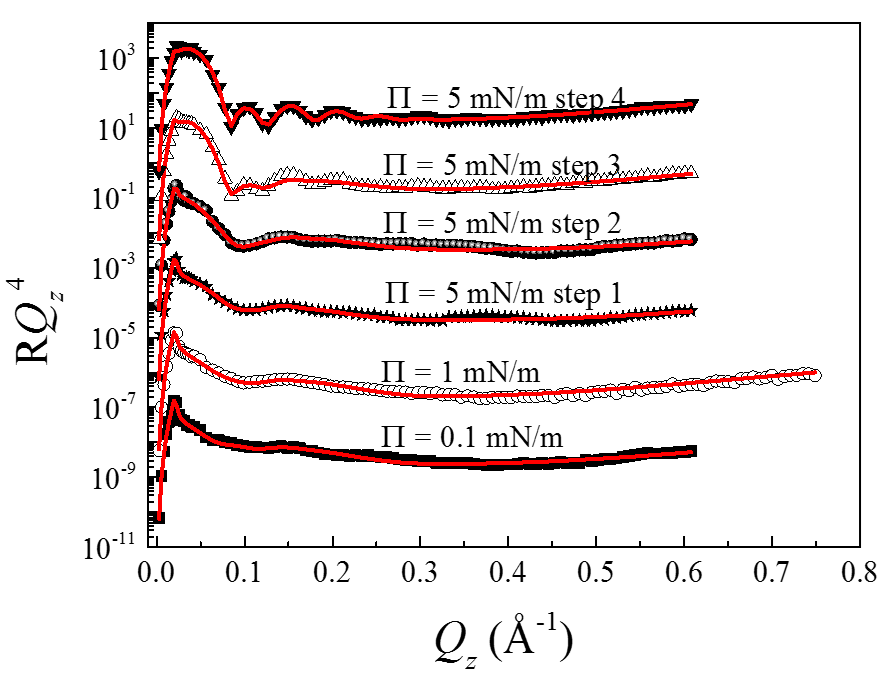}
\caption{XRR experimental data (symbols) and fitted models (solid lines) for BM10$_3$20 sample at pressures $\Pi=0.1$ mN/m, $\Pi=1$ mN/m and $\Pi=5$ mN/m. The later one is repeated four times in a sequential order.}
\label{Fig3}
\end{figure}

Transverse structure of the layer was examined by means of XRR, which provides information on scattering length density (SLD) as a function of a distance $z$ from the top interface with air ($z = 0$). Reflected intensity collected with a linear PSD was background-corrected and fitted to a model SLD profile using Parratt formalism \cite{12} and least-squares fit analysis. In general, deviation of the experimental reflectivity curve from Fresnel reflectivity (smooth decay of intensity proportional to $Q_z^4$) elicits a layering process occurring at sample surface or interface. The argument $Q_z=\frac{4\pi}{\lambda} sin(\alpha_i)$ is a component of momentum transfer vector perpendicular to a surface, where $\alpha_i$ is an angle of incidence angle and $\lambda$ is photon wavelength. As reflectivity function $R(Q_z)$ has an asymptotic decay $Q_z^{-4}$, our experimental XRR curves (Fig.\ref{Fig1} and Fig.\ref{Fig3}) are presented in form of $RQ_z^4$ to give an access to the smallest features of the data, and to emphasize quality of the fit.

Experiments on 10 nm MNPs were performed at room temperature and at pressure ranging from 5 to 10 mN/m. Upon reaching the pressure $\Pi = 10$ mN/m several XRR curves were measured every 5 minutes in a process of relaxation of the layers with the trough barrier activated (denoted step 1, step 2, etc. in Fig. \ref{Fig1}).  Figure \ref{Fig1} shows experimental XRR data (symbols) with the fitted model curves (solid lines). Corresponding in-depth SLD profiles are shown in Fig. \ref{Fig2}. Upon increasing the pressure the thicknesses of the layer remains equal to the size of a single nanoparticle, while its SLD increases indicating progressive densification of the layer.

The XRR curves obtained from MD20 in the pressure range from $\Pi = 1$ mN/m to 19 mN/m at room temperature demonstrate only Fresnel decay of the intensity without any modulations. Thus, the XRR experiment gives an evidence that the 20 nm nanoparticles do not create neither a homogeneous layer nor large enough lateral clusters. Specifically, for our experimental setup only clusters of size of a few microns and larger could be detected by the XRR technique which is related to a coherence length of the photon beam.

\begin{figure}
\includegraphics[width=8cm]{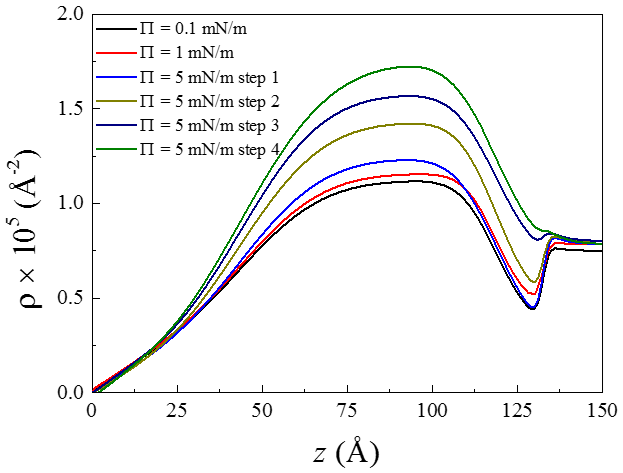}
\caption{SLD profiles of BM10$_3$20 at different steps of monolayer formation at pressure values from $\Pi=0.1$ mN/m to $\Pi=5$ mN/m.}
\label{Fig4}
\end{figure}

XRR measurements on BM10$_3$20 sample were also carried at room temperature. Pressure in the film was varied from 0.1 mN/m to 5 mN/m. Figure \ref{Fig3} shows the XRR experimental data (symbols) with corresponding best fit models (lines). Similar to the case of MD10 sample one can see increase of the SLD value upon the compression (Fig. \ref{Fig4}). This increase is more pronounced in a part of the layer adjacent to the water surface ($75 < z < 130$~\AA). Surprisingly, the total thickness of layer is almost the same as for the monodisperse sample MD10. Thus, it can be assumed that shape of the XRR curves for the BM10$_3$20 sample is determined mostly by the 10 nm nanoparticles while the 20 nm particles are undetectable by the reflectivity method in the same manner as it was observed for the monodisperse sample MD20. This means that the 20 nm particles are not embedded into the layer.

\subsection{Grazing Incident Small-Angle X-ray Scattering}

\begin{figure*}
\includegraphics[width=16cm]{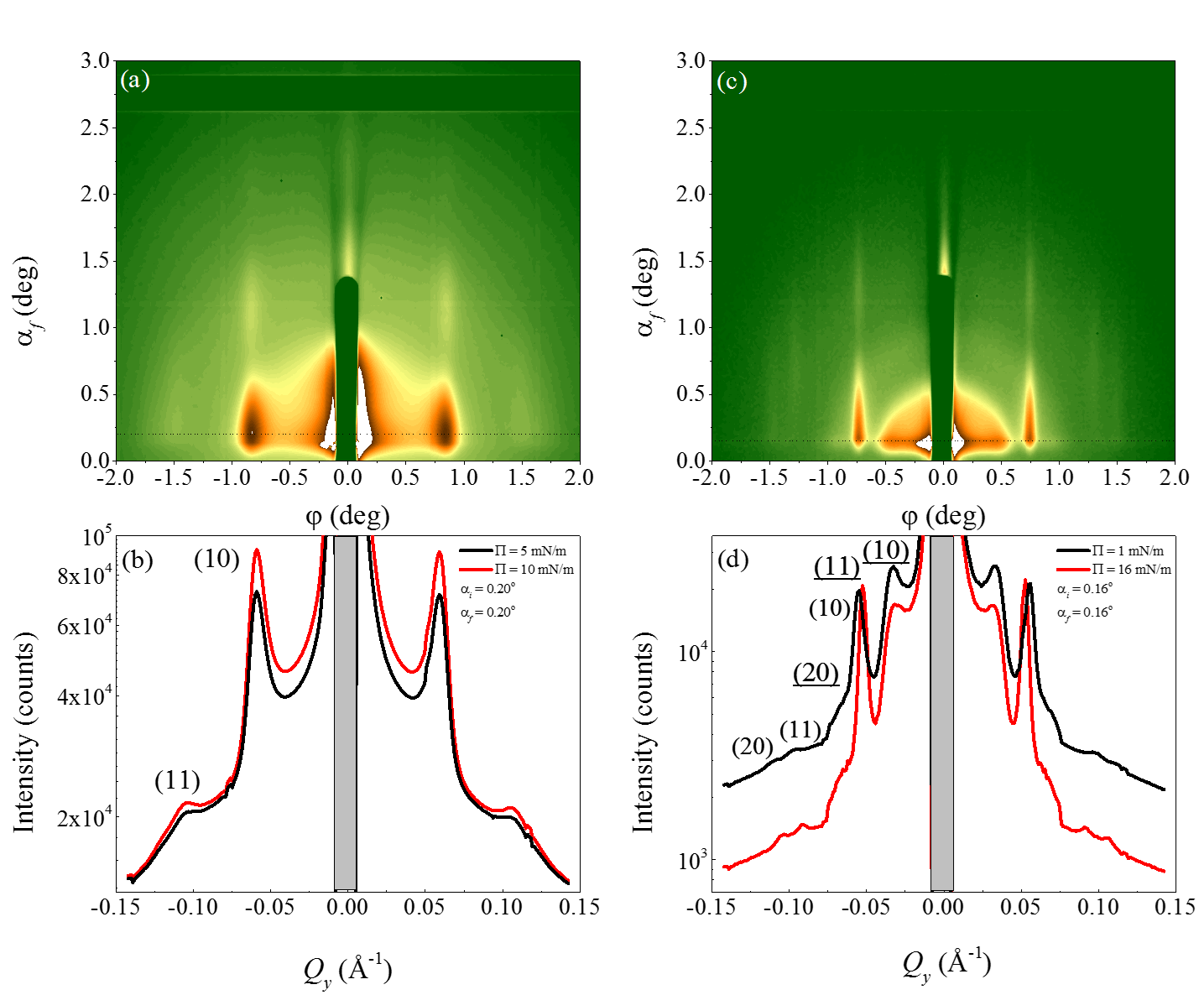}
\caption{GISAXS scattering patterns at maximum pressure in the layers. (a) Two-dimensional pattern for MD10 at $\Pi = 10$ mN/m taken at $\alpha_i = 0.20^\circ$. (b) Intensity distribution along the line $\alpha_f=0.2^\circ$ for MD10 at $\Pi = 5$ mN/m and $\Pi = 10$ mN/m. (c) Two-dimensional pattern for BM10$_3$20 at $\Pi = 16$ mN/m taken at $\alpha_i = 0.16^\circ$. (d) Intensity distribution along the line $\alpha_f=0.16^\circ$ for BM10$_3$20 at $\Pi = 1$ mN/m and $\Pi = 16$ mN/m.}
\label{Fig6}
\end{figure*}

Fig.\ref{Fig6} shows the GISAXS data obtained from MD10 and BM10$_3$20 samples at low and high pressure respectively. The Bragg peaks, originating from an in-plane ordering of the particles, are extended along the $\alpha_f$ direction manifesting two-dimensional nature of the samples. To determine accurately the peaks positions $Q_{y}^{hk}$, appropriate cuts across the two-dimensional scattering patterns were taken. Position of the cut is shown by a dashed line.

For the MD10 sample at both pressure values $\Pi = 5$ mN/m and $\Pi = 10$ mN/m (Fig.\ref{Fig6}a,b) only (10) and (11) peaks corresponding to a hexagonal lattice with a constant $a = 12.4$ nm are observed. Corresponding calculated and measured interplanar distances $d^c_{hk}$ and $d^m_{hk}$ are presented in Table \ref{tab}.

The GISAXS data for the BM10$_3$20 sample are shown in Fig.\ref{Fig6}c,d. Two sets of the Bragg peaks can be distinguished. They correspond to two hexagonal lattices with constants $a_1 = 13.2$ nm, $a_2 = 22.5$ nm at $\Pi = 1$ mN/m and $a_1 = 13.8$ nm, $a_2 = 22.7$ nm at $\Pi = 16$ mN/m (see Table \ref{tab}). Bragg peak indexes corresponding to the subsystems with $a_1$ and $a_2$ are shown in Fig.\ref{Fig6}, whereas those for the subsystem with $a_2$ are underlined.

\begin{table*}
\caption{\label{tab} Comparison of calculated interplanar distances $d^c_{hk} = \frac{a}{\sqrt{\frac{4}{3}(h^2+hk+k^2)}}$ calculated assuming a hexagonal lattice with a constant $a$ with measured interplanar distances $d^m_{hk} = \frac{2\pi}{Q_y^{hk}}$ for the monodisperse sample at pressures $\Pi=5$ mN/m and $\Pi=10$ mN/m (in brakes) and for the binary sample at pressure $\Pi=1$ mN/m and $\Pi=16$ mN/m (in brakes). }
\begin{center}
\begin{tabular*}{\textwidth}{ccccccc}
\hline\hline
Sample &$h$& $k$& $d^c_{hk}$, nm& $a_{1,2}$, nm & $d^m_{hk}$, nm&\\
\hline
MD10 &1& 0 & 10.7 (10.7) & 12.4 (12.4) &10.7 (10.7)\\
 & 1 & 1 &6.2 (6.2) & & 6.2 (6.2) \\
\hline
BM10$_3$20 (10 nm MNPs) &1& 0 & 11.4 (12.0) & 13.2 (13.8) &11.4 (12.0)\\
 & 1 & 1 & 6.6 (7.0) & & 6.6 (6.9)\\
 & 2 & 0 & 5.8 (5.9) & & 5.8 (5.9)\\
\hline
BM10$_3$20 (20 nm MNPs)  &1& 0 & 19.5 (19.7) & 22.5 (22.7)&19.5 (19.7)\\
 & 1 & 1 & 11.3 (11.4) & & 11.4 (12.0) \\
 & 2 & 0 & 9.8 (9.8) & & 9.5 (9.7) \\
    \hline\hline
\end{tabular*}
\end{center}
\end{table*}

\subsection{Scanning Electron Microscopy}

\begin{figure}
\includegraphics[width=8cm]{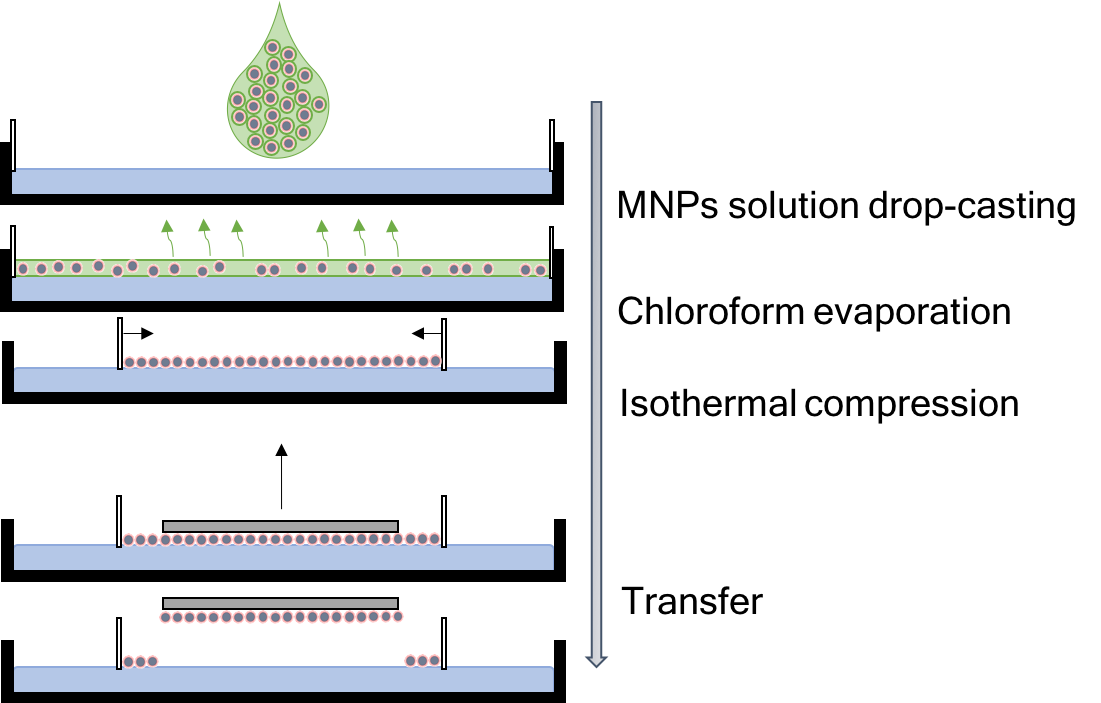}
\caption{Sketch of the Langmuir-Schaefer deposition technique.}
\label{FigLS}
\end{figure}

The samples were deposited on a solid substrate of lateral size $20\times 20$ mm using Langmuir-Schaefer technique (stamping) after assembling in a Langmuir through (Fig. \ref{FigLS}). As a substrate we used silicon wafer coated with layer of gold, which was functionalized with a layer of 1-pentadecanethiol molecules making it hydrophobic for better adhesion of the MNPs. The details of MD10 and MD20 assembling, choice of the substrate and deposition routine are discussed elsewhere \cite{A,ukleev2016x}. 

\begin{figure*}
\includegraphics[width=17cm]{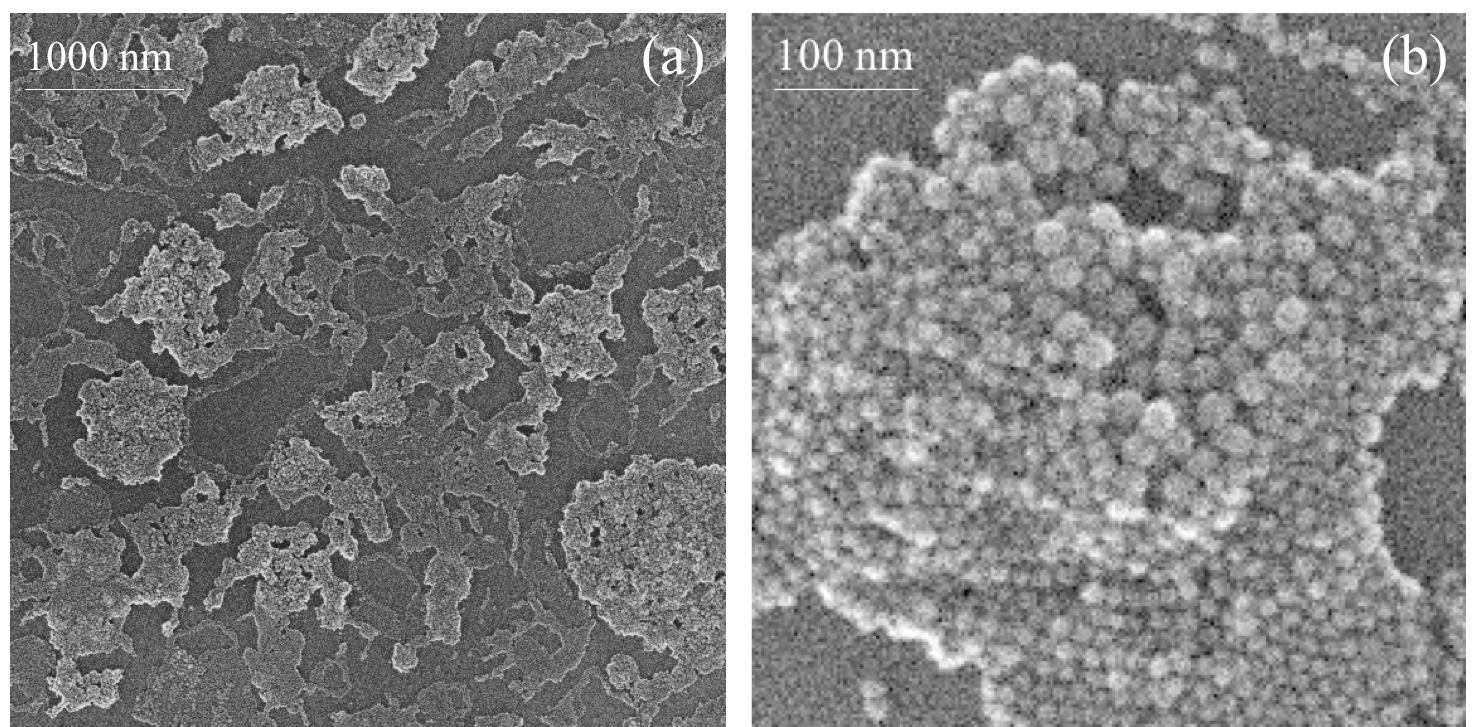}
\caption{SEM images of the BM10$_3$20 sample deposited on a solid substrate with different magnifications.}
\label{Fig11}
\end{figure*}

Figure \ref{Fig11}a,b shows  BM10$_3$20 sample deposited onto a substrate at $\Pi = 16$ mN/m using Langmuir-Schaefer technique. One can clearly observe the monolayer of 10 nm particles and three-dimensional agglomerates of 20 nm MNPs. Such agglomerates of the large particles were always observed on the top of the monolayer of the small particles and never on the bare substrate. Analysis of the SEM images gives mean value of the interparticle distance equal to $12.8 \pm 0.7$ nm, which is in a good agreement with the GISAXS results.

\section{Discussion}

Results of the XRR experiment demonstrate that 10 nm particles in the binary mixture BM10$_3$20 form a homogeneous monolayer which density increases upon the compression. The larger particles were not detected by the reflectivity measurements what could indicate absence of those particles on the water surface due to, for instance, sinking of them down. However, the GISAXS patterns show unambiguously that both types of the particles are present close to the water surface. Therefore, one should infer that 20 nm particles form small, sub-micrometer, clusters which are undetectable by XRR.

Analysis of the GISAXS data confirms periodic lateral structure of the BM10$_3$20 layer. The layer contains two different subsystems, each assembled from the particles of one size -- either 10 nm or 20 nm. The subsystem consisting of 10 nm particles has two-dimensional hexagonal structure. At low pressure (1 mN/m) the lattice constant $a_1$ of this structure is equal to 13.2 nm, that is significantly larger than corresponding lattice constant in the monodisperse sample MD10 -- 12.4 nm. Moreover, it increases to 13.8 nm with the pressure rise to 16 mN/m. In contrast, in MD10 the value of the lattice constant does not depend on pressure. Intensity of the Bragg peaks increases with rise of pressure what indicating overall densification of the 10 nm system. This observation is consistent with the increase of the peak SLD density observed in the XRR measurements. Interestingly, the width of the peaks for BM10$_3$20 is much more narrow as compared to that for MD10. Therefore, two close peaks (11) and (20) can be resolved only for BM10$_3$20. According to the Scherrer formula \cite{smilgies2009scherrer}, this means that lateral coherence length (or average size of a hexagonally ordered domains) of 2-D crystal assembled from 10 nm particles is much larger in the binary sample. Formation of larger domains is probably manifested also in low-sloped shape of the BM10$_3$20 isotherm corresponding to the LC state of the sample (Fig. \ref{Fig0}). This implies that particles have more time to optimize their arraignment (slow growth process). Increase of the lattice constant $a_1$ with rise of pressure is not fully understood. One can suggest that the underlying clusters of 20 nm particles stretch  hcp lattice of 10 nm particles by means of specific magnetic interaction.

\begin{figure}
\includegraphics[width=8.5cm]{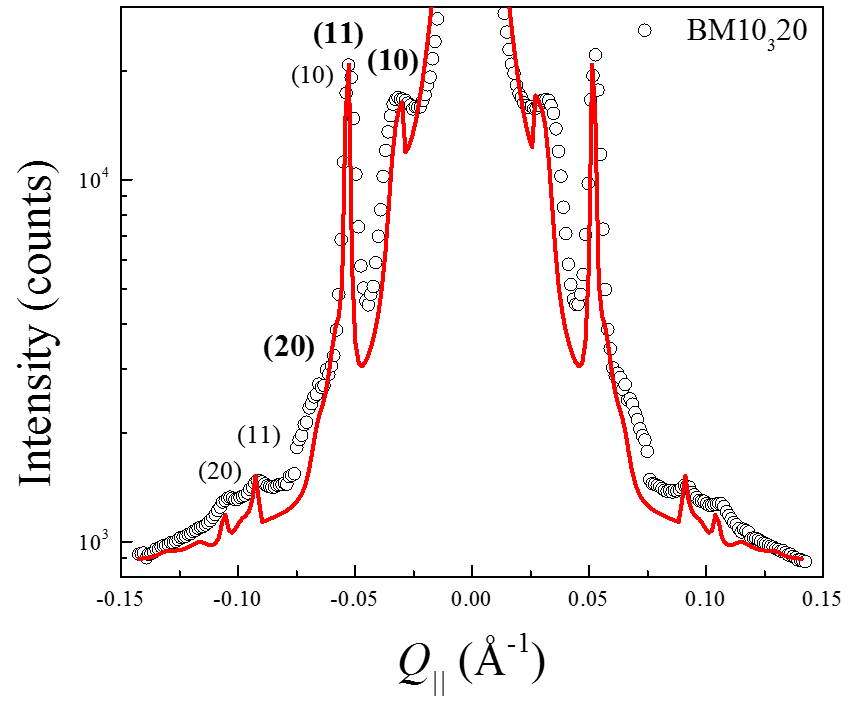}
\caption{Comparison of GISAXS pattern calculated as incoherent sum of contributions from two hexagonal lattices assembled from 10 nm and 20 nm particles (line) with the experimental data obtained from BM10$_3$20 at $\Pi=16$ mN/m. The lattice parameters used in calculations are shown in Table \ref{tab}.}
\label{Fig10}
\end{figure}

For the second subsystem consisting of 20 nm particles only one diffraction peak (10) can be reliably identified on the GISAXS patterns. Higher order peaks of supposedly hexagonal structure are barely discernible. Peak (11) coincides with peak (10) of the 10 nm subsystem. Next low-intensity peak (20) would be situated in a region of sharp rise of background intensity and could be seen only as a small hump. However, model simulations of GISAXS pattern from arrays of spherical nanoparticle ordered in hcp lattices performed in BornAgain package \cite{16} confirm presence of those peaks and reveal their positions (see Fig.\ref{Fig10} and Table \ref{tab}). Consequently, the lattice constant $a_2$ can be obtained. In the calculations it was assumed that hexagonal subsystems of spherical 10 nm and 20 nm MNPs contribute to the scattering patterns incoherently. Good fit of the model curve to the experimental data affirms formation of two types of hexagonal monodisperse clusters each consisting either of 10 nm particles or of 20 nm particles. This result, however, does not explain expansion of the lattice lattice constant of 10 nm MNPs subsystem in binary mixture.

In our previous study \cite{A} it was shown that monodisperse 10 nm iron oxide particles create ordered clusters of size of at least a few microns immediately after spreading over the water surface. The clusters do not interact and do not change their size unless they brought in to contact with each other during the compression. This explains extremely extended LE+G region of the isotherms. In contrast, monodisperse 20 nm particles form sub-micrometer clusters with a strong long-range interaction between them. As a result, the isotherm starts straight away from the LC state although each particle occupies an area which could accommodate several tens of such particles. The same behavior was observed in the present study for the samples MD10 and MD20 (Fig.\ref{Fig0}). Interestingly, the isotherm for the binary sample BM10$_3$20 looks like a superposition of the BM10 and BM20 isotherms. This can be assumed as an indirect proof of size-selective separation obtained by XRR and GISAXS which is caused, most probably, by the peculiarities of the dipole-dipole interaction in this system. 

There is a strong evidence that magnetic dipole-dipole interaction plays important role in self-organization of MNPs \cite{A}. When energy of this long-range interaction becomes comparable with the thermal energy, the system of MNPs undergoes a transition from a superparamagnetic to a collective state. Although blocking temperature of superparamagnetic ensembles of iron oxide MNPs of size about 9 nm is reported around 125 K \cite{pichon2011tunable, pauly2012size}, a monolayer of 20 nm particles can be in a ferromagnetic state already at room temperature \cite{santoyo2011magnetic, carvalho2013iron}. It should be also pointed out that the effect of dipolar interactions is stronger in thin films compared to liquid or powder samples as a consequence of the shape anisotropy \cite{pauly2012size}. Dipolar interaction tend to align magnetic moments of nanoparticles along local magnetic field (or stray field) to increase attraction (reducing potential energy) of the ensemble. Parallel orientation of the dipoles leads to attractive interaction, while antiparallel arrangement of the same dipoles produces repulsion between them, and the magnitude of the interaction for parallel alignment is more energy favorable \cite{singamaneni2011magnetic}. This interaction is completely described by the dipole-dipole representation, as the stray field caused by the single homogeneously magnetized nanoparticle is equivalent to a point dipole field due to the spherical symmetry:

\begin{equation}
\begin{split}
\mathbf{H}_{dip} (\boldsymbol{\mu}_j, \mathbf{r}_{i,j}) = \frac{3 \mathbf{r}_{i,j} (\boldsymbol{\mu}_j  \mathbf{r}_{i,j}) - \boldsymbol{\mu}_j r_{i,j}^2}{r_{i,j}^5},
\end{split}
\end{equation}
where $\mathbf{H}_{dip}$ is magnetic field,  $\mu = \frac{4}{3} \pi M_s R^3$ is a magnetic moment (R is particle radius, $M_s$ is saturation magnetization),$\mathbf{r}_{i,j}$ is distance between $i$ and $j$ dipoles. Therefore the energy of dipole-dipole interaction between two particles can be written as: 
\begin{equation}
\begin{split}
U_{dip}(\mathbf{\mu_j}, \mathbf{r}_{i,j}) = -\boldsymbol{\mu}_i \mathbf{H}_{dip} (\boldsymbol{\mu}_j, \mathbf{r}_{i,j}).
\end{split}
\end{equation}

There are several other factors to be taken into account when self-assembly of nanoparticles on water surface is considered. The most important of them are Van der Waals and steric interactions and surface tension. The van der Waals interaction of the London type refined by Hamaker \cite{hamaker1937london} is attractive:

\begin{equation}
\begin{split}
{U_{VdW}} = - \frac{A}{6} (\frac{2}{l^2+4l}+\frac{2}{(l+2)^2}+ln\frac{l^2+4l}{(l+2)^2}).
\end{split}
\end{equation}

Here $l = s/R$, $s$ is distance between sphere surfaces, $A \approx 10^{-19}$ is the Hamaker constant \cite{fertman1990magnetic}.
Steric interaction is determined by the the surfactant-solvent environment. In our case it is oleic acid in water and air. Since water is a polar molecular liquid, there is the steric attraction of surfactant molecules at length-scale $\sigma \approx 2$ nm (thickness of the surfactant shell). It is difficult to estimate the order of this force, but it is clear that the electrostatic forces on such the distance are dominant \cite{fertman1990magnetic}. The surface tension has microscopic nature and becomes significant only for the large particles or agglomerations on water surface ($\approx 5$ microns or larger) \cite{kralchevsky2000capillary}. Once the clusters are formed, the surface tension can trigger their further coalescence.

\begin{figure}
\includegraphics[width=8cm]{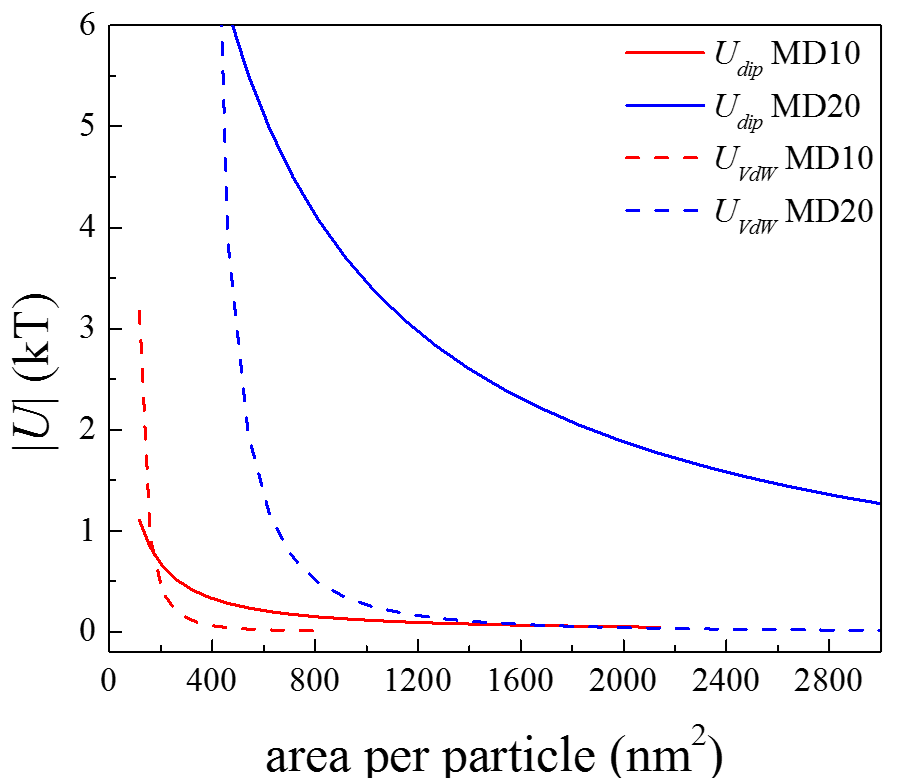}
\caption{Magnetic dipole-dipole and Van der Waals interaction energies for two nanoparticles of type MD10 and MD20 calculated according to Eq-s. (2) and (3).}
\label{Fig14}
\end{figure}

Therefore for our systems MD10 and MD20 only dipolar and Van der Waals energies can be considered. Estimated values of those energies are presented in Fig. \ref{Fig14} as functions of the area-per-particle parameter to be directly compared to the isotherms shown in Fig.\ref{Fig0}. As one can see, long-range dipole-dipole interaction between MNPs dominates in a wide range of the area-per-particle values, but for 10 nm particles it is much smaller. Since large particles magnetically interact at longer distances, one can presume that exactly dipole-dipole interaction leads them to form clusters only with each other ignoring smaller particles at early stages of the layer formation.

In order to gain another insight into the origin of size-separation we performed micromagnetic modeling of magnetization distribution in binary and single-component MNP clusters. Although only magnetic contribution to the whole MNPs interaction energy was taken into account, results of calculations appear to be consistent with the experimental results. Since thermal effects were not considered (as well as magnetocrystalline anisotropy), our model is more adequate for the case of low temperature. However, we assume that qualitatively it should give identical results even at room temperature. 

The exchange coupling constant $B = 10^{-11}$ J/m  and the saturation magnetization $M_s = 3.8 \cdot 10^5$ A/m used in calculations correspond to bulk maghemite \cite{thanh2012magnetic, dunlop2001rock}. It was shown recently that saturation magnetization of iron oxide MNPs is rather low at room temperature and, moreover, depends on the particle size in a non-trivial way defined by an actual core-shell structure of the particles \cite{gabbasov2015mossbauer}. In our calculation a simple approximation for the magnetic moment $\mu = \frac{4}{3} \pi M_s R^3$ was used. Correspondingly, ratio between magnetic moments of 20 nm and 10 nm MNPs is equal to 8, with is quite high. However, in reality it can be even higher \cite{gabbasov2015mossbauer}. 

The calculations are based on Landau-–Lifshitz-–Gilbert equation which can be written as:
\begin{equation}
\frac{d \mathbf{M}}{d t} = -\gamma' \mathbf{M} \times \mathbf{H}_{\mathrm{eff}} + \frac{\alpha}{M_s} \mathbf{M} \times \frac{d \mathbf{M}}{d t},
\end{equation}
where $\mathbf{M}$ is the magnetization vector which depends on the space coordinates and time, $\gamma$ is the Gilbert gyromagnetic ratio, $\alpha$ is the phenomenological damping parameter, $\mathbf{H}_{\mathrm{eff}}$ is the effective field. Since we are interested only in the final metastable state we take the damping parameter equal to 1. 

We used Nmag micromagnetic simulation package \cite{fischbacher2007systematic} provided by University of Southampton. Hybrid Finite Element/Boundary Element method which is implemented in Nmag is highly effective for multiple bodies calculations. It makes it possible to build mesh only in bodies but not in the whole space. Maximum linear size of finite element did not exceed 1.2 nm and, hence, it was smaller than maghemite exchange length $l_{ex} = \sqrt{2 B / (\mu_0 M_s^2)} \approx 10$ nm. While finite element calculation is rather computationally challenging it guarantees correct evaluation of magnetic state inside of the nanospheres and interactions between the nanospheres. 

\begin{figure}
\includegraphics[width=8cm]{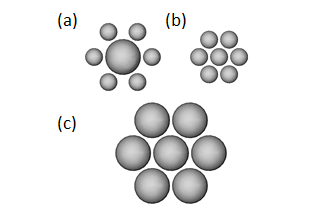}
\caption{Structure of small clusters of MNPs examined by means of micromagnetic modeling: (a) binary cluster, (b) monodisperse MD10 cluster, (c) monodisperse MD20 cluster.}
\label{Fig15}
\end{figure}

Demagnetization energy and magnetization configuration for different monodisperse and binary cluster were calculated in saturation remanence state. Obtained remanence magnetization configurations were vortex-like for all considered clusters. It is known that this state corresponds to deep energy minimum \cite{bennett2003simulating,baskin2012clusters}. Although discussion of magnetic states of the MNPs clusters is out of scope of the present paper, some of typical configurations can be found in Supplementary materials. Here we focus on comparison of total demagnetization energies of several MNPs clusters which can form in the binary sample BM10$_3$20. 

We started with small clusters consisted of seven close packed nanospheres (Fig. \ref{Fig15}). Diameter of small and large nanospheres was 9.3 and 18.6 nm correspondingly. Distance between nanospheres surfaces was set to 1.1 nm to take into account nonmagnetic coating layer. Our main goal was to analyze the demagnetization energies of monodisperse and binary clusters. Therefore we compared total demagnetization energy of monodisperse clusters consisted of small and large nanospheres (Fig. \ref{Fig15}b,c) with the energy of binary clusters (Fig. \ref{Fig15}a). The easiest way to do this is to keep the number of large and small nanospheres equal. Since binary cluster consists of one large and six small nanospheres we multiplied its demagnetization energy by a factor of 7. After that we could directly compare this value with total energy of one cluster made of large nanospheres and six clusters consisted of small nanospheres. Here we assumed that clusters do not interact with each other. It was found that the total energy of monodisperse clusters (4.78 eV) is lower than that of binary clusters (5.39 eV). Therefore it can be concluded that formation of small monodisperse clusters is energetically more favorable. In other words small nanospheres do not stick to single large ones. 

\begin{figure}
\includegraphics[width=8cm]{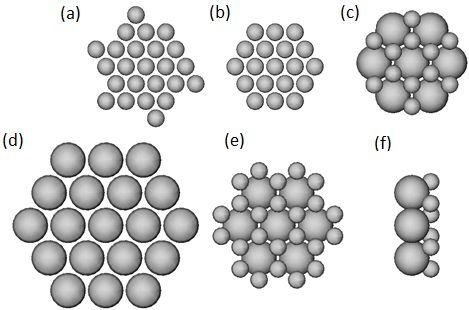}
\caption{Large monodisperse (a), (b), (d) and binary (c), (e) clusters that were studied. (f) is the side view of the cluster (c).}
\label{Fig16}
\end{figure}

The situation changes with increasing of the size of the clusters as it follows from consideration of close packed highly symmetrical clusters presented in Fig. \ref{Fig16}. We constructed binary clusters shown in Fig. \ref{Fig16}c,e taking into account experimental GISAXS data suggesting that both small and large nanospheres form hexagonal lattices. Moreover, in such assembly the lattice constant 13.3 nm of the small nanospheres agrees with corresponding experimental value 13.2 nm if we introduce measured value of lattice constant for large spheres cluster -- 22.5 nm. Finally, stoichiometric composition of binary mixture leads to definite structure of the binary cluster.

\begin{figure*}
\includegraphics[width=17cm]{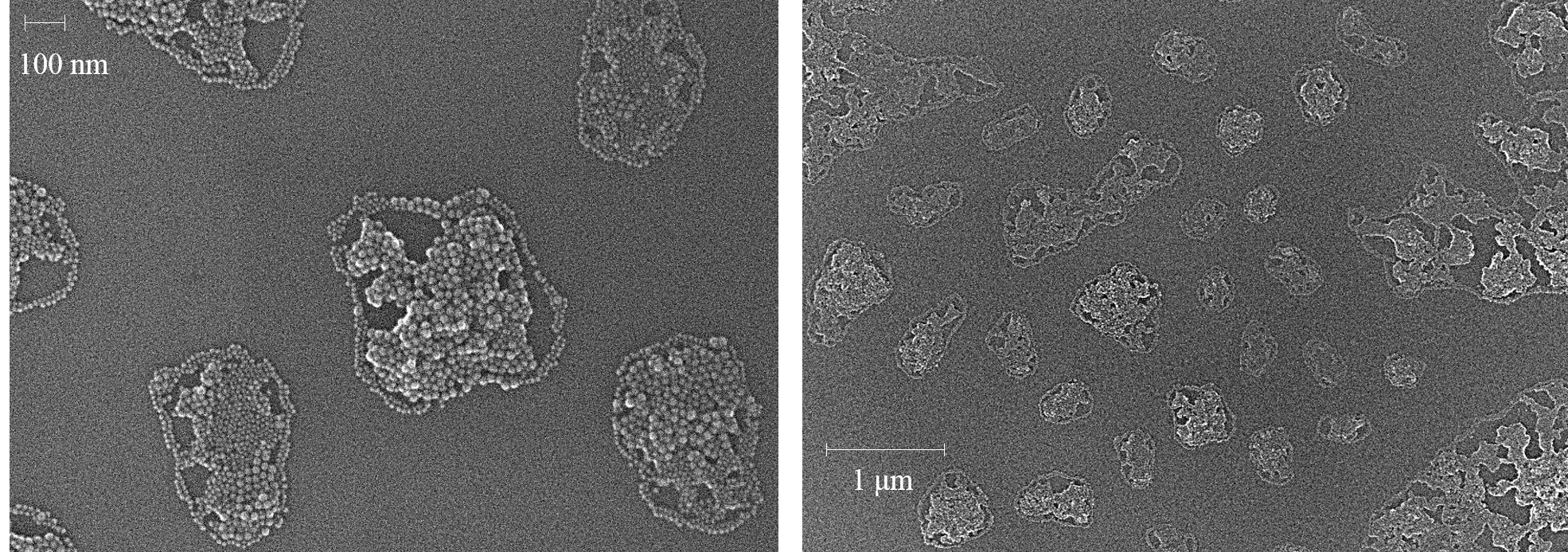}
\caption{SEM image of the BM10$_3$20 sample annealed on the subphase by infrared heater for 12 minutes and deposited on a solid substrate by the Langmuir-Schaefer technique.}
\label{Fig17}
\end{figure*}

Using above-described procedure we examined all options of combining large and small nanospheres into monodisperse and binary clusters presented in Fig. \ref{Fig16}. We found out that in all cases demagnetization energy of binary clusters is about 5\% lower than that of monodisperse ones. 

Therefore, analysis of pure magnetic interactions suggests that at initial stage MNPs create small monodisperse islands. However, after those clusters adhere to each other and they can form structures shown in Fig. \ref{Fig16}c,e,f. It should be noted that energy difference per particle between monodisperse and binary clusters is of the order of $0.5kT$ ($T = 293$ K), hence, other factors like Van der Waals interactions can also affect MNPs arrangement. 

The experimental methods, used in the present study, do not allow us to elicit reliably how the clusters of different types are arranged in respect to each other. It would be reasonable to assume, that all clusters lie in one horizontal plane. More long-range ordering of 20 nm nanoparticles in the binary sample than in the monodisperse sample (where only one Bragg peak was observed by GISAXS) gives an evidence in support of this assumption. However, on SEM image shown in Fig.\ref{Fig17} one can clear see a cluster of 20 nm particles overlying a cluster of 10 nm particles. Assuming, that such constructions form already on the liquid surface, and taking into account, that the deposition was done using the Langmuir-Schaefer method (stamping, turning the film upside down), one could conclude, that in the original Langmuir film the aggregates 20 nm particles underlie the monolayer formed by the 10 nm particles. Thus, for the cluster shown in Fig.\ref{Fig16}f its left side would be adjacent to the water surface. However, it can not be guaranteed, that the clusters of 20 nm particles have not changed their positions during or after the deposition. This question will be addressed in our future work.

To provide different ordering of the binary mixture we attempted to anneal the BM10$_3$20 sample directly on water surface. For annealing the infrared heater of ID10 beamline for in-situ heating experiments was used. Sample was heated to the temperature about 60C for 12 minutes, then cooled down for 30 minutes and deposited to the solid substrate. However, as it can be seen in Fig. \ref{Fig17}, annealing is only stimulating nanoparticles separation. Langmuir film is divided by monolayer islands of 10 nm MNPs of 0.1 -- 2 $\mu m$ size, within 20 nm MNPs clusters inside. Thus the most likely scenario for nanoparticle organization is corresponding to Fig. \ref{Fig16}c,e,f which is supported by enlargement of lattice constant of 10 nm MNPs monolayer in case of BM10$_3$20 mixture. The average crystalline domain size is corresponding to the cluster size shown in Fig. \ref{Fig17}. However, the origin of the cluster separation after annealing process is unclear. We can speculate, that long-range arrangement was diminished upon heating, while the short-range ordering was protected by magnetic dipolar and Van der Waals interactions. More detailed investigation of the influence of the annealing process to Langmuir films of iron oxide MNPs will be discussed elsewhere.

Recently, J. Stanley et al. \cite{stanley2015spontaneous} used the similar X-ray scattering methods to study the ordering of binary mixtures of 10 nm and 20 nm iron oxide nanoparticles with different concentration ratios. They obtained similar experimental results as the present work and came to conclusion about phase separation between MNPs subsystems by size. However, the small change of the lattice constant of 10 nm MNPs subsystem was not discussed.

\section{Conclusion}

Experimental XRR and GISAXS data accomplished with model calculations and SEM shows that introduction of iron oxide MNPs of two different sizes, 10 nm and 20 nm, in one system does not lead to the formation of a two-dimensional binary crystal in the Langmuir film. The subsystem of 20 nm iron oxide MNPs forms a three-dimensional structures, while 10 nm MNPs are organized in the monolayer. Addition of the 20 nm MNPs to ensemble leads to the enlargement of the lattice constant of 10 nm MNPs subsystem, what hints at non-trivial mixing of nanoparticles of two sizes. We assume that magnetic dipole-dipole interaction is the most powerful driving force of the self-assembly process. Once the array of 20 nm MNPs are formed, the surface tension promotes their further aggregation and simultaneous drowning resulting in three-dimensional clusters. Further formation of monolayer of 10 nm MNPs is influenced by this assemblies. 

\section*{Acknowledgments}
Authors thanks European Synchrotron Radiation Facility for the provided beamtime and technical assistance. Scanning electron microscopy research was performed at the Research park of St.Petersburg State University Interdisciplinary Center for Nanotechnology and European Synchrotron Radiation Facility. This work was supported by the Russian Foundation for Basic Research, grants 12-02-12066-ofi m and 14-22-01113-ofi m. M. Dudnik thanks A. Eliseev (Moscow State University) for a fruitful discussion. 

\bibliographystyle{model1a-num-names}
\bibliography{biblio}
\end{document}